# Numerical Study of Two-Frequency Heating in ECR Ion Source

*E.Yu. Kidanova [a,]*, V.E. Mironov [a], D.K. Pugachev [a]*

*[a]Joint Institute for Nuclear Research, Flerov Laboratory of Nuclear Reactions, Dubna, Moscow Region, 141980, Russia*

**kidanova@jinr.ru*

**Abstract.** The two-frequency heating (TFH) method is widely used to improve the performance of electron cyclotron resonance (ECR) ion sources, yet its physical mechanisms remain insufficiently understood. In this work, the TFH regime in the DECRIS-PM ECR ion source is studied numerically using the NAM-ECRIS code. Single-frequency heating at 14.51 GHz (500 W) is compared to two-frequency heating at 14.51 GHz and 13 GHz (450 W + 50 W) at equal total power. It is found that while the global plasma parameters remain largely unchanged, the peak density of highly anisotropic electrons ($E > 1$ keV, $V_\perp/V_\parallel > 10$) decreases by ~20% and their distribution broadens radially in the TFH regime. This reduction in electron anisotropy is attributed to the randomizing effect of the lower-frequency wave on electron gyromotion, which suppresses plasma micro-instabilities and leads to improved ion production.

**Keywords:** Electron cyclotron resonance ion source; ECRIS plasma; Two-Frequency Heating; Numerical Simulation; Electron Heating

## 1. Introduction

Electron cyclotron resonance (ECR) ion sources are the primary tool for producing intense beams of highly charged ions for nuclear physics accelerator facilities [1]. The operating principle of an ECR ion source is based on the resonant energy transfer from a microwave wave to plasma electrons in a magnetic field: when the electron cyclotron frequency matches the microwave frequency, electrons are efficiently heated at the resonance surface.

Several technical methods allow substantial improvement of source performance without radical changes to its design: wall coatings, gas mixing, biased electrode, and multiple frequency heating [2]. Among these, the two-frequency heating (TFH) method is one of the most effective.

The TFH method consists of the combined injection of two microwave waves at different frequencies into the plasma chamber. The method was successfully applied in 1994 by the Lawrence Berkeley National Laboratory (LBNL) group on the AECR source using two klystrons at 10 and 14 GHz [3]. A shift of the charge state distribution toward higher values was observed, beam intensities increased by factors of 2–5, plasma stability improved, and greater total microwave power could be coupled.

Despite the extensive experimental basis, the theoretical understanding of TFH mechanisms remains the subject of active investigation. Initially, the TFH effect was explained by the enlargement of the effective microwave power absorption volume due to the appearance of a second resonance surface, leading to increased heating stochasticity and higher average electron temperature [2]. The INFN-LNS group [8] performed a series of numerical simulations within the single-particle approach and argued that TFH increases the electron heating rate, deepens the plasma potential dip, and thereby increases the ion confinement time and extracted beam current. The FAR-TECH group [9] identified in numerical simulations an energy barrier in the single-frequency regime: electrons moving along momentum-space curves determined by a given frequency reach a limiting energy and cannot be heated further. TFH allows this barrier to be overcome by randomizing the electron movement in momentum space, which results in a higher plasma temperature.

Experimentally, it was demonstrated in [10] that TFH suppresses the electron cyclotron instability in ECR ion source plasma. The authors of [10] conjecture that the hot electron scattering into loss-cone by the second-frequency waves reduce the anisotropic electron density inside the plasma. The possibility is also mentioned that the hot electron component is locally redistributed with TFH inside the magnetic trap by shifting the reflection points of electron bounce oscillations due to formation of inner secondary ECR surface. Subsequent studies [11] experimentally confirmed that the TFH regime suppresses instabilities in the ECR plasma. It was suggested that two-frequency regime enhances the parallel diffusion component in the velocity space, which reduces the anisotropy of the electron energy distribution function and thereby decreases the instability growth rate.

In the present work, a numerical study of the two-frequency heating regime is carried out with an aim to provide an interpretation of the TFH effect.

## 2. General description

The numerical experiment was performed using the NAM-ECRIS code [12], a three-dimensional code for simulation of plasma in ECR ion sources. The code is based on the particle-in-cell with Monte Carlo collisions (PIC-MCC) method. A detailed description of the model is given in [12], below we outline its key components.

The code includes two PIC modules for electrons and ions that work in an iterative loop. The microwave field distribution is computed every iteration step using the RF module of COMSOL Multiphysics in the cold-plasma approximation and is then passed to the electron module as a fixed input.

The electron module receives the ion parameters and the COMSOL field distributions. It treats the electron interaction with the microwave field, electron-ion and electron-electron collisions, and ionization energy losses. Anomalous Bohm-like diffusion is included in regions of radially decreasing magnetic field by adding random kicks perpendicular to the field line. The kick frequency is a free parameter; it is set to 1 MHz, a value at which the simulation results are not sensitive to its exact choice. Electron interactions with the chamber walls are modeled using the Furman-Pivi secondary electron emission (SEE) scheme.

Electrons move in electric and magnetic fields simultaneously: the static magnetic field of the trap, the oscillating microwave field, and the sheath electric field near the walls. The sheath is characterized by the plasma potential, taken as a free parameter, and by an enlarged sheath width of 2.5 mm in the axial direction and 1 mm near the radial walls. The plasma potential creates a barrier that retains cold electrons and controls their lifetime. Its value is chosen to match the experimentally measured extracted ion currents. An additional sheath enhancement is applied near the injection-side electrode, which is biased at −250 V.

The calculations were carried out for the DECRIS-PM ion source [13]. Two heating regimes were simulated for argon plasma: single-frequency heating (SFH) at 14.51 GHz with 500 W, and two-frequency heating (TFH) at 14.51 GHz + 13 GHz with 450 W + 50 W. The total power is the same in both cases, allowing a direct comparison.

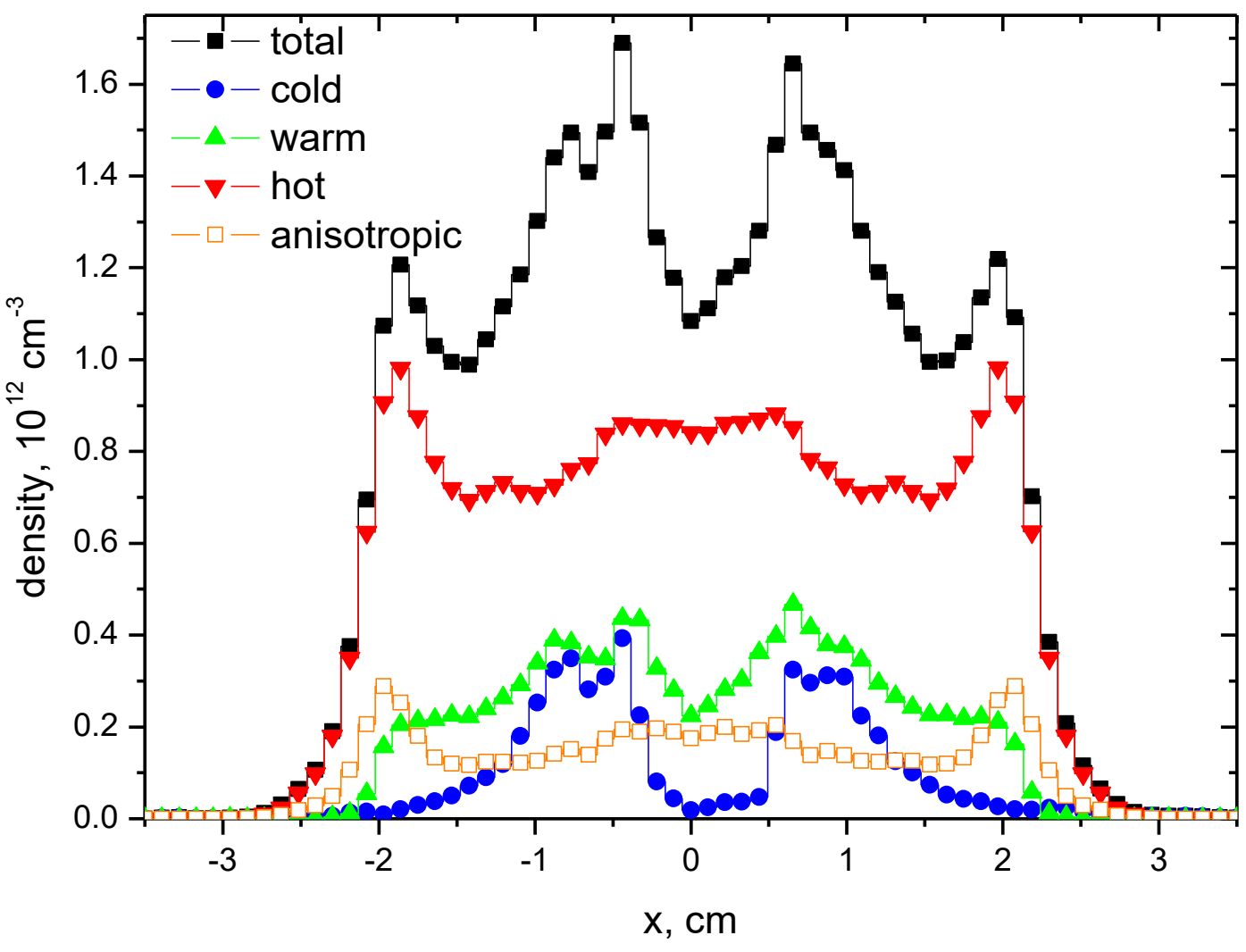

Fig. 1. – Transversal profile of electron densities at the center of the source: black - total density; blue - cold electrons with energy $E_e$ up to 0.25 keV, green - warm electrons with $E_e$ from 0.25 to 5 keV; red – hot electrons ($E_e \geq 5$ keV), orange – anisotropic electrons, $E_e \geq 1$ keV and $V_\perp/V_\parallel \geq 10$.

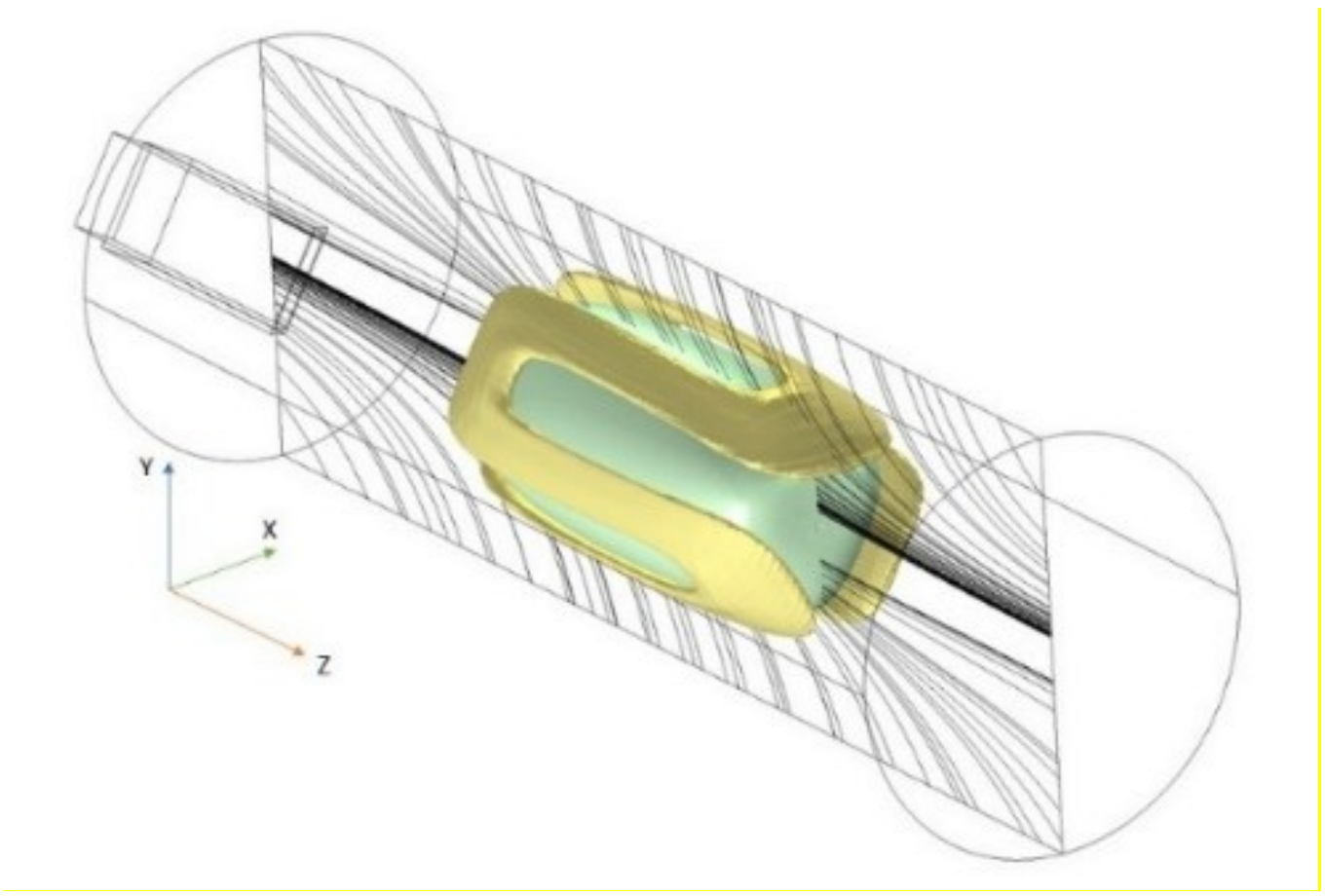

Fig.2. – Regions with the highest density of the anisotropic electrons ($E_e \geq 1$ keV and $V_\perp/V_\parallel \geq 10$). The ECR zone is shown as the green ellipsoid, the magnetic field lines are shown as the black lines.

The calculated electron density profiles for SFH mode are shown in Fig. 1. The transversal density profile is shown at the axial position where magnetic field is minimal, close the source chamber center. The peak total density reaches approximately $1.7 \times 10^{12}$ cm$^{-3}$, the profile is is concentrated inside the ECR surface. The total density is plotted in black, and the contributions from cold (0–0.25 keV, blue), warm (0.25–5 keV, green), and hot (above 5 keV, red) electrons are shown separately. In addition to partitioning the plasma electrons into the cold, warm and hot components, we also select those electrons that have energies above 1 keV and the ratio between the perpendicular and longitudinal velocities in respect to the local magnetic field ($V_\perp/V_\parallel$) above 10, designating them as the anisotropic electrons. These electrons represent the population, which is supposed to be responsible for the instability excitation in ECR plasma [10]. In Fig.1, their transversal profile is shown as orange open squares.

The three-dimensional spatial distribution of the anisotropic electrons is illustrated by Fig.2, where regions are shown where the anisotropic electron density is above 70% of its maximal $3\times10^{11}$ cm$^{-3}$ value. Also, the ECR zone for 14.51 GHz waves is shown in green, as well as the magnetic field lines in the y-z plane. The anisotropic electrons are preferentially localized close outside the ECR zone in the regions where the magnetic field lines approach tangentially the zone surface.

### 3. Comparison of the plasma parameters in the single- and two-frequency heating modes.

The major visible impact of the TFH is observed in the maximal density and spatial distribution of the highly anisotropic electrons in plasma. In Fig.3, the spatial distributions of the anisotropic electron density are shown in the transversal x-y plane close to the minimum-B position along z-axis for the single and double frequency heating modes (a and b), and in the x-z plane (c and d), with the plane orientation shown in Fig.2. The ECR zone boundaries $B_{res1}$=0.52 T and $B_{res2}$=0.46 T are shown in black for 14.51 and 13 GHz microwaves respectively.

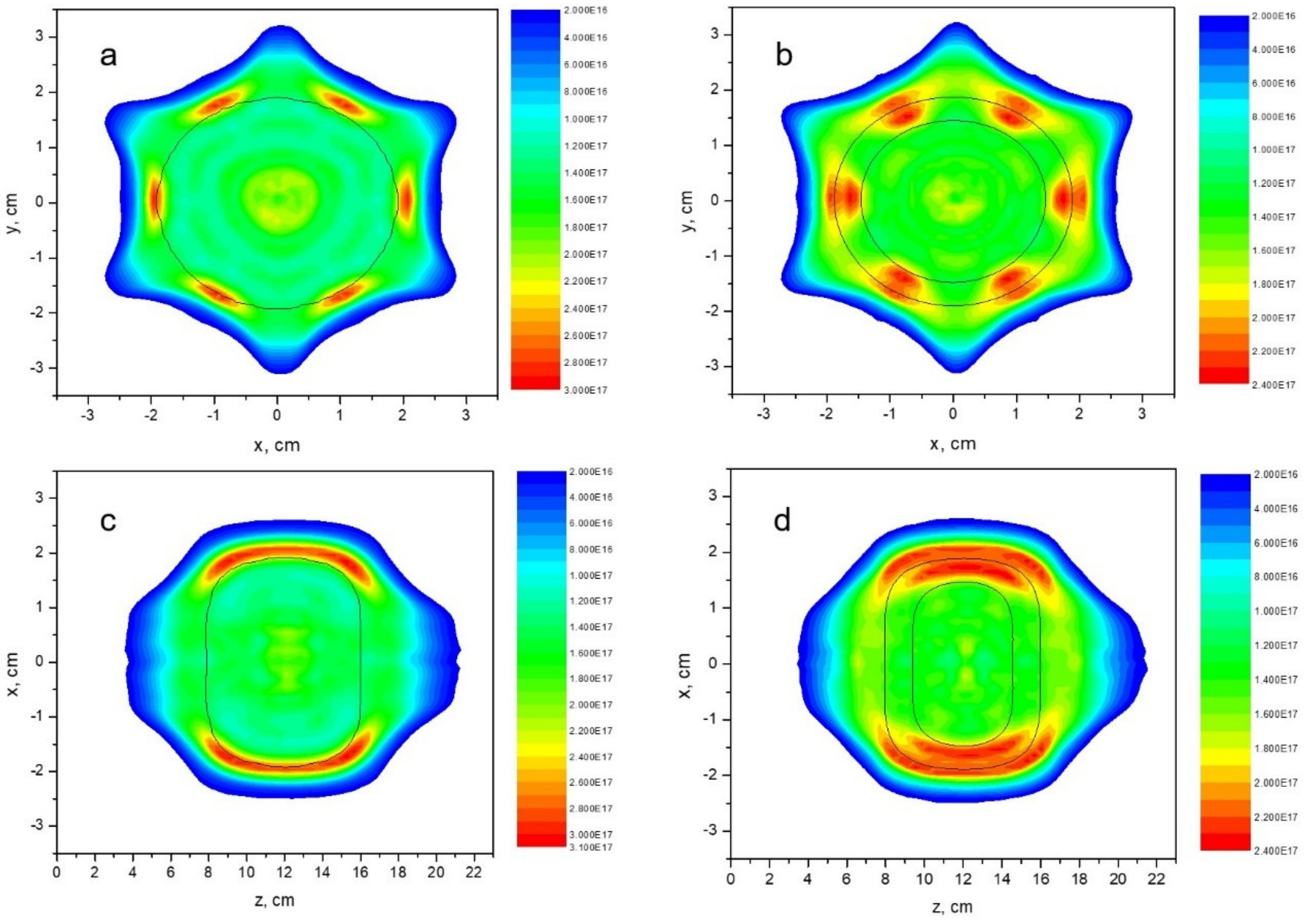

Fig.3. – Spatial distributions of anisotropic electron density in the transversal x-y plane for SFH mode (a, $n_{emax}$=3.0×10$^{11}$ cm$^{-3}$) and TFH mode (b, 2.4×10$^{11}$ cm$^{-3}$), and in the longitudinal x-z plane for SFH (c, 3.1×10$^{11}$ cm$^{-3}$) and TFH (d, 2.4×10$^{11}$ cm$^{-3}$).

The transversal profiles of anisotropic electron density are shown in Fig.4 at the same longitudinal position as in Fig.3ab along y=0.0 line in SFH (red) and TFH (black) modes. Drops in the maximal values and broadening of the distributions are seen with injection of the lower frequency microwaves.

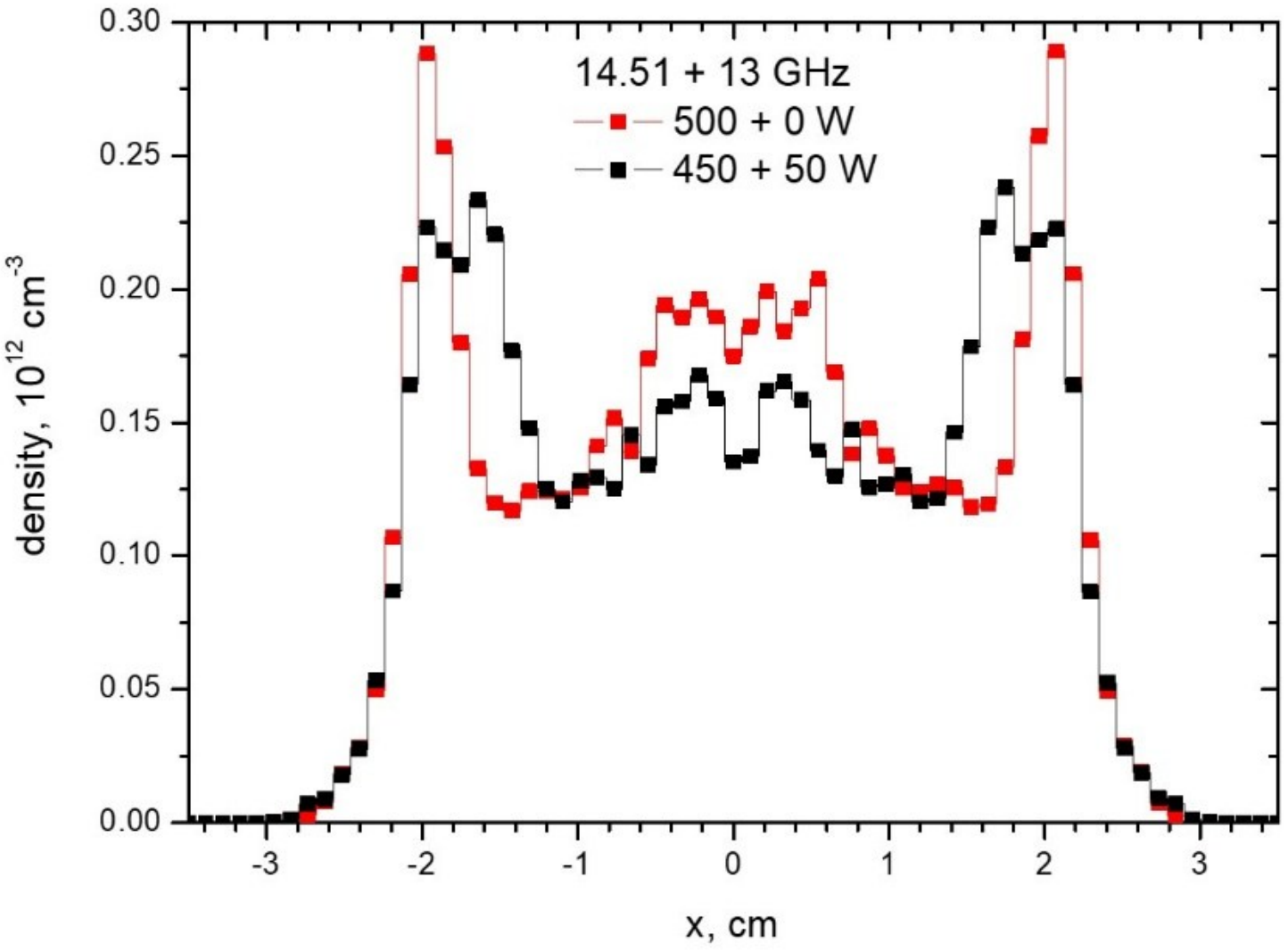

Fig. 4. – Comparison of the density of anisotropic electrons in plasma for SFH (red) and TFH (black) modes.

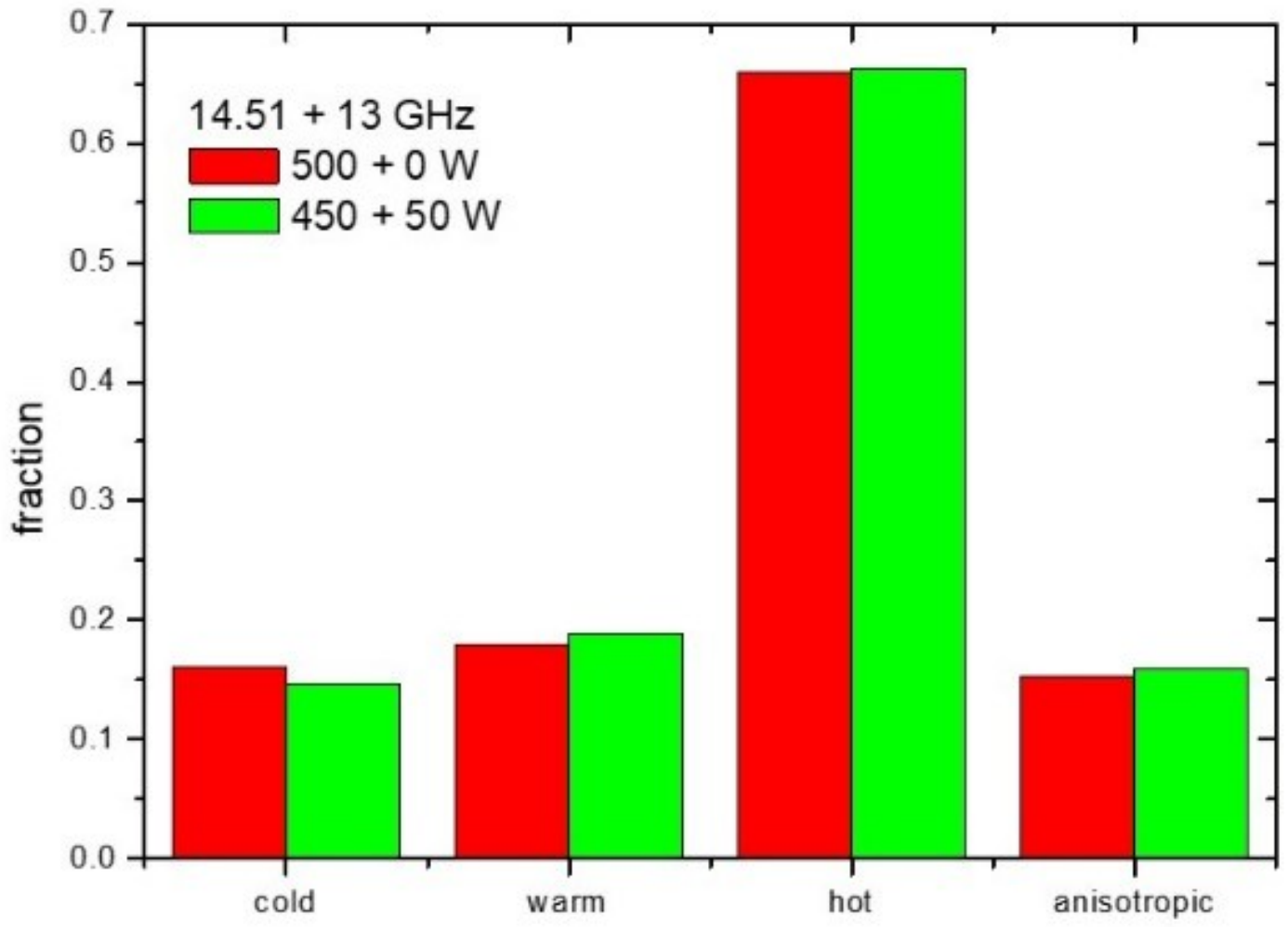

Fig. 5. – Fractions of cold, warm, hot and anisotropic electron components relative to the total electron number for the SFH (red, 500 + 0 W) and TFH (green, 450 + 50 W) regimes.

In the SFH regime, the anisotropic electron population is strongly localized close outside the ECR surface, forming a well-defined narrow structure in the x-y plane. In the TFH, this distribution broadens radially, and the peak density decreases by 20%. The total electron density distributions are nearly identical in both modes, showing no noticeable changes in the shape or magnitudes. Similarly, the electron energy spectra exhibit no significant difference between the SFH and TFH cases, indicating that the introduction of the second frequency at 13 GHz with 50 W does not alter the global plasma parameters - the total electron density and the overall energy content of the plasma remain essentially unchanged. This is confirmed by the electron population fractions shown in Fig. 5. The fractions of the cold, warm, hot, and anisotropic electron components are the same in both regimes.

## 4. Discussion

Our calculations do not support those explanations of the two-frequency heating effect that involve variations in electron energy distributions or in a content of the highly anisotropic electrons. We see a decrease in the anisotropic electron densities connected to re-distribution of these electrons in a space, with essentially the same content in the plasma. Drop in the density can influence excitation rates of a plasma instability, allowing quiet plasma generation and more effective production of the highly charged ions in the source.

We look now into details of how the secondary frequency influences parameters of the anisotropic electrons in ECRIS plasma. For this, additional diagnostics is applied: when advancing particle's velocities with the pusher, average differences are calculated in energy and in angle $\Phi$ between the longitudinal and perpendicular velocities in respect to the magnetic field direction (for the numerical reasons we calculate the complementary value to the pitch angle). For the non-relativistic electrons, the variations are fully determined by the local amplitude of the microwave electric field and by the phase between the electron gyromotion and the field oscillations. The results are shown in Fig.6 for electrons with energies below 1 keV for whole population and for the anisotropic electrons as a function of the magnetic field at the particle advance position. The fields are normalized by the ECR value for 14.51 GHz, the resonant field for 13 GHz waves is 0.9 in these units. The angle variations are plotted with changing their sign to represent changes in the pitch angle.

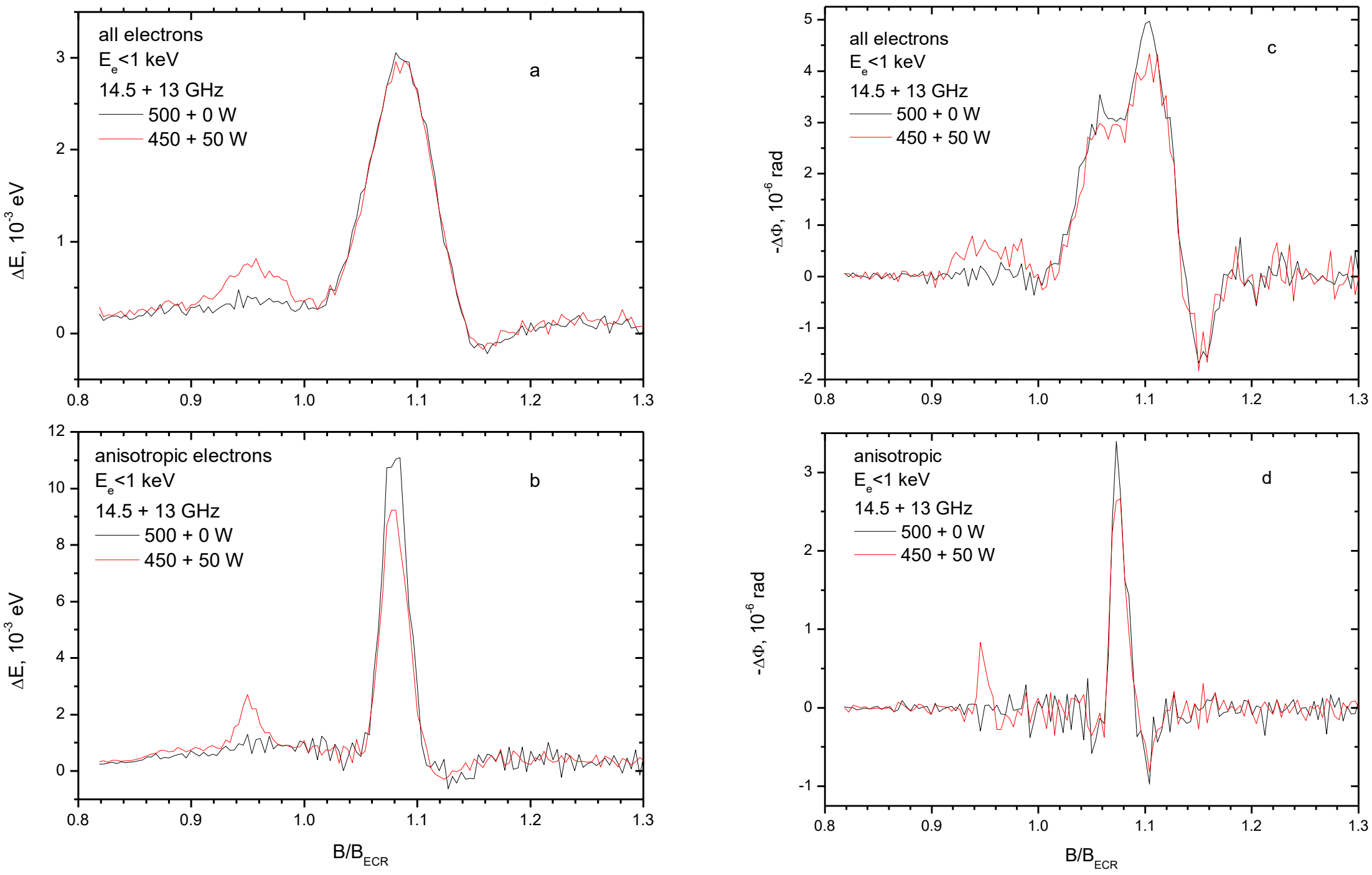

Fig.6 – Average electron energy changes in one timestep as a function of the local magnetic field for all (a) and anisotropic electrons (b); average changes in the angle (-ΔΦ) between the longitudinal and perpendicular velocity components for all (c) and anisotropic electrons (d). The black lines represent the SFH mode, the red lines are for the TFH mode.

In the SFH mode, electrons are heated inside the Doppler-broaden resonance layer adjacent to the "cold" ($\omega_{rf}=\omega_{ce}$) ECR surface with $B/B_{ECR}$=1.0. In the double-frequency heating mode, additional layer appears in the regions close to the low-frequency resonance $B/B_{ECR}$=0.9. The heating rate for all electrons with $E_e$<1 keV by the main microwaves is not changing significantly when switching to the TFH mode. For the anisotropic electrons in the SFH mode, heating profile is narrower due to the reduced Doppler broadening, and the peak value is increased by factor of ~3. Strong reduction in the peak value of heating rate by the main-frequency microwaves is observed in the TFH mode. Basically, the same dynamics is observed for the pitch-angle variations: only a small difference between two heating modes is seen in the changes calculated for all electrons with $E_e$<1 keV, while for the anisotropic electrons the peak value is reduced, and the additional peak appears in the regions at the second resonance layer.

If electrons interact with the microwaves always with random phase in respect to the electric field oscillations, the positive and negative kicks in the perpendicular velocities cancel each other and the averaged pitch angle variations should be zero. Graphs c) and d) in Fig.6 show that electron heating process in ECRIS is not fully stochastic as it is often assumed in the ECR heating models. The spatial distribution of anisotropic electrons shown in Fig.2 indicates that substantial amount of these electrons is mirror-trapped inside or close to the resonance layer, while moving perpendicular to the magnetic field lines due to the curvature drift. Phase and frequency of their gyromotion is synchronized with the microwaves causing continuous increase of their perpendicular velocities. The addition of the 13 GHz wave disrupts this phase coherence, scattering the anisotropic electrons away from the ECR surface and reducing their peak density. We note that potentially the same decoherence effect can be achieved by a frequency of phase modulation of the primary microwaves. This is planned to be a subject of our further numerical and experimental investigations.

### CrediT authorship contribution statement

**Ekaterina Kidanova:** Writing – original draft, Software, Investigation, Visualization. **Vladimir Mironov:** Writing – original draft, Software, Investigation, Conceptualization, Visualization. **Dmitriy Pugachev:** Conceptualization, Writing – review & editing, Supervision.

### Declaration of competing interest

The authors declare that they have no known competing financial interests or personal relationships that could have appeared to influence the work reported in this paper.

### Data availability

Data will be made available on request.

**Funding sources**

This research did not receive any specific grant from funding agencies in the public, commercial, or not-for-profit sectors.